# Strain Tunable Berry Curvature Dipole, Orbital Magnetization and Nonlinear Hall Effect in WSe$_2$ Monolayer

Mao-Sen Qin, Peng-Fei Zhu, Xing-Guo Ye, Wen-Zheng Xu, Zhen-Hao Song, Jing Liang, Kaihui Liu, Zhi-Min Liao*

State Key Laboratory for Mesoscopic Physics and Frontiers Science Center for Nano-optoelectronics, School of Physics, Peking University, Beijing 100871, China.

*Email: liaozm@pku.edu.cn

**Abstract:** **The electronic topology is generally related to the Berry curvature, which can induce the anomalous Hall effect in time-reversal symmetry breaking systems. Intrinsic monolayer transition metal dichalcogenides possesses two nonequivalent K and K' valleys, having Berry curvatures with opposite signs, and thus vanishing anomalous Hall effect in this system. Here we report the experimental realization of asymmetrical distribution of Berry curvature in a single valley in monolayer WSe$_2$ through applying uniaxial strain to break C$_{3v}$ symmetry. As a result, although the Berry curvature itself is still opposite in K and K' valleys, the two valleys would contribute equally to nonzero Berry curvature dipole. Upon applying electric field *E*, the emergent Berry curvature dipole *D* would lead to an out-of-plane orbital magnetization $M \propto D \cdot E$, which further induces an anomalous Hall effect with a linear response to $E^2$, known as nonlinear Hall effect. We show the strain modulated transport properties of nonlinear Hall effect in monolayer WSe$_2$ with moderate hole-doping by gating. The second-harmonic Hall signals show quadratic dependence on electric field, and the corresponding orbital magnetization per current density $M/J$ can reach as large as 60. In contrast to the conventional Rashba-Edelstein effect with in-plane spin polarization, such current-induced orbital magnetization is along the out-of-plane direction, thus promising for high-efficient electrical switching of perpendicular magnetization.**

The Berry curvature, which describes the local geometry of Bloch wavefunction,[1] plays an essential role in modern condensed matter physics. In the view of charge transport in crystals, Berry curvature can produce an anomalous velocity,[2,3] which is related to various exotic topological phenomena, including anomalous Hall effect,[4] valley Hall effect,[5] anomalous Nernst effect[6] and the negative magnetoresistance in topological materials.[7] In addition, the integral of Berry curvature in parameter space can give rise to the topological Chern invariant,[1] which is one of the major subjects in topological systems. Generally, to obtain nonzero Berry curvature in a system, either time-reversal symmetry or inversion symmetry needs to be broken.[8] Recently, a new class of electromagnetic response, related to the Berry curvature in the nonlinear regime, has been revealed based on both semiclassical approach and quantum theory, including circular photogalvanic and nonlinear Hall effect.[9–16] It is proposed that the dipole moment of Berry curvature in momentum space, *i.e.*, Berry curvature dipole, should be responsible for such nonlinear behaviors.[9] In contrast to the conventional Hall effect, the nonlinear Hall effect is generated as a second-order response to external electric field $E$ without the involvement of external magnetic field. The nonlinear Hall current $J_H$ is associated with the Berry curvature dipole $D$ by $J_H \propto (D \cdot E)\hat{z} \times E$, where the term $D \cdot E$ induces orbital magnetization $M$ and an anomalous Hall effect.[17–19] The Berry curvature dipole has been proposed and experimentally confirmed in two-dimensional $WTe_2$, manifested by both the nonlinear Hall and circular photogalvanic effect.[17–19] Furthermore, the ferroelectric switching of Berry curvature dipole in multi odd-layer $WTe_2$ was proposed and observed, which may be applicable for non-volatile memory based on Berry curvature.[20,21]

The transition metal dichalcogenide (TMD) monolayers with $1H$ structure are ideal platforms to discovery the Berry curvature-related physics, including the Berry curvature dipole induced nonlinear Hall effect.[11] The $1H$ structure TMDs with hexagonal crystals possess two nonequivalent $K$ and $K'$ valleys in Brillouin zone, promising for the applications in the field of valleytronics.[5,22] The inversion symmetry is spontaneously broken in TMD monolayers accompanied with strong spin-orbit coupling,[11] which produces a large Berry curvature. Opposite valleys hold the Berry curvatures with opposite signs, leading to valley contrasting physics,[5] such as the valley optical selection rule, valley Hall effect and valley Zeeman splitting.[6,23–27] However, the presence of $C_{3v}$ symmetry in monolayer TMDs forces the Berry curvature dipole to be zero, leading to the vanishing nonlinear Hall effect.[11] In fact, the largest symmetry, which is allowed by the nonzero Berry curvature dipole, is a single mirror symmetry in two-dimensional materials.[9,11] Thus, to observe nonlinear Hall effect, the $C_{3v}$ symmetry needs to be broken, for example, by means of uniaxial strain, in which case only a mirror symmetry is preserved.[11] Indeed, the strained TMDs are extensively studied by optical measurements previously, where it is found that strain can induce bandgap decreasing and valley magnetization.[28–30] Nevertheless, the strain in TMDs is usually applied through the organic flexible substrates, hindering the systemic investigations about the strain engineering of transport properties.

In this work, we report the electrical-tunable Berry curvature dipole and nonlinear Hall effect in strained monolayer $WSe_2$. The uniaxial strain is applied by the single crystal $(1-x)[Pb(Mg_{1/3}Nb_{2/3})O_3]$-$x[PbTiO_3]$ (PMN-PT) piezoelectric substrate. The PMN-PT is with high piezoelectric coefficient and high dielectric constant.[31–34] By controlling the strain values, the system shows a highly tunable Berry curvature dipole, and thus a controllable out-of-plane orbital magnetization under an external electric field.

The monolayer $WSe_2$ flakes were mechanically exfoliated from bulk crystals onto $SiO_2$/Si substrates, which were pre-cleaned in air plasma for 1 minute. To apply strain in the desired direction, the flakes with long, straight edges (which were generally along the crystalline axis, such as zigzag or armchair direction[18]) were selected and subsequently transferred onto the PMN-PT substrate based on the standard dry transfer technique. The long, straight edges of the flakes were aligned with the [001] orientation of PMN-PT crystal. The piezoelectric field $E_P$ was applied on the PMN-PT crystal along the [001] orientation to induce strain along the same direction. The samples were patterned into standard Hall bars by air plasma etching, and the Ti/Au electrodes were prepared by electron beam lithography, electron beam evaporation of metal and lift-off. The transport measurements were carried out in a commercial Oxford cryostat. The first- and second-harmonic signals were collected by standard lock-in techniques with frequency $\omega = 17.777$ Hz.

Figure 1(a) shows the schematic of the WSe$_2$ device. In addition to the conventional Hall bar structure, a voltage $V_P$ was applied on the PMN-PT substrate to generate the piezoelectric field $E_P$ along the [001] direction with the distance between the two electrodes of 50 μm. The ionic-liquid gating voltage ($V_g$) was applied to tune the Fermi level into the valence band and realize high hole conduction.[35] The atomic force microscope image of a typical device is shown in Fig. 1(b), where the Hall bar is denoted by purple. The crystalline direction is identified by the polarization-resolved second harmonic generation (SHG). Figure 1(c) shows the second harmonic intensity component parallel to the polarization as a function of the laser polarization angle, where the directions with maximum intensity correspond to the armchair direction of WSe$_2$.[30] By the SHG measurements, it is found that the applied $E_P$ direction is along the zigzag direction with negligible misalignment in the device of Fig. 1(b).

Upon applying piezoelectric field $E_P$ along the zigzag direction, strain is induced along the same direction in WSe$_2$. Such uniaxial strain may also possess significant effect on the band structure, such as bandgap engineering and valley shift.[36–39] Here we only focus on the strain induced nonzero Berry curvature dipole as shown in Figs. 1(d-f). The expression for Berry curvature dipole can be written as[11,18]

$$D_\alpha = -\frac{1}{\hbar} \int \delta(\varepsilon - \varepsilon_F) \frac{\partial \varepsilon}{\partial k_\alpha} \Omega(k) d^2 \mathbf{k},$$

where $\varepsilon$ is the energy, $k_\alpha$ is the wavevector along $\alpha$, $\varepsilon_F$ is the Fermi energy and $\Omega$ is the Berry curvature. The $\delta$ function indicates the integral is nonzero only near the Fermi energy. The integral is in fact determined by two parts, *i.e.*, the band slope $\frac{\partial \varepsilon}{\partial k_\alpha}$ and Berry curvature $\Omega(k)$. For unstrained TMDs shown in Fig. 1(d), the time-reversal symmetry requires the Berry curvature to be opposite on opposite valleys.[23] For a single valley (K or K'), the Berry curvature is symmetric while $\frac{\partial \varepsilon}{\partial k_\alpha}$ is antisymmetric, inducing zero Berry curvature dipole (bottom panel in Fig. 1(d)). Therefore, to obtain nonzero Berry curvature dipole, Berry curvature must possess asymmetric distribution in a single valley. Such asymmetric distribution is able to be induced by uniaxial strain,[29] as shown in Figs. 1(e) and 1(f) for tensile and compressive strain along the zigzag direction, respectively. For tensile strain in Fig. 1(e), the Berry curvature extrema is shifted away from the vertex of $K$ /$K'$ valleys,[11] leading to nonvanishing Berry curvature dipole along the zigzag direction (bottom panel in Fig. 1(e)). For compressive strain in Fig. 1(f), opposite process would happen, inducing opposite Berry curvature dipole. Intriguingly, as shown in Figs. 1(e-f), in a single valley, the Berry curvature dipole generally forms between the minimum and maximum of Berry curvature, and the two valleys contribute equally to the Berry curvature dipole due to the protection of time-reversal symmetry.

To confirm the strain induced by the PMN-PT substrate, the resistance of WSe$_2$ as a function of $E_p$ was investigated at 140 K and is shown in Fig. 2(a). It is worth noting that multiple sweeps of $E_p$ are required before the strain-driven resistance change stabilize into a standard hysteretic loop. For direct comparison, we also fabricated strain gauge patterned using the Ti/Au zigzag strips in the piezoelectric field region. The gauge resistance as a function of $E_P$ at 140 K is shown in Fig. 2(b), demonstrating similar hysteretic behaviors. It is worth noting that the piezoelectric polarization cannot be switched to opposite direction after poling the PMN-PT in our experiments because the applied $E_P$ is not enough compared to the large coercivity. [40] That is to say, the strain varies monotonously without sign change by sweeping the piezoelectric field from positive to negative.[34] During the sweep process, we first apply positive piezoelectric field up to $20\,\mathrm{kV/cm}$, at which the gauge resistance is increased, indicating the positive piezoelectric field corresponds to tensile strain. Then the piezoelectric field is decreased towards $-20\,\mathrm{kV/cm}$, leading to the suppression of tensile strain.

The strain level $\varepsilon$ can be roughly estimated by $\frac{\Delta R}{R}/K$, where $\frac{\Delta R}{R}$ is the resistance change by strain and $K$ is the gauge factor. For the Au gauge, $K$ is close to 2.[34]

According to Fig. 2(b), the strain level at 140 K is estimated to be around 0.2%, 0.1% and nearly zero for $E_P = $ 15, 0 and -15 kV/cm, respectively. However, the size of the Au gauge (around 100 μm × 100 μm) is much larger the individual domain size of the piezoelectric substrate. Thus, the strain levels estimated by Au gauge is an average effect, while small WSe$_2$ flake may feel domain-dependent strain.[34] As calculated from Fig. 2(a), the change of resistance, defined as $\frac{\Delta R}{R_{min}} = \frac{R_{max}-R_{min}}{R_{min}} \times 100\%$ as sweeping the $E_P$, is ~110% for WSe$_2$. If using the gauge factor of TMDs as large as 40,[41] the strain level in WSe$_2$ can be roughly estimated to be around 2.8% for $E_P = 15$ kV/cm. However, the strain induced change of band structure can significantly influence the resistance change in WSe$_2$ channel, making the accurate calculation of the gauge factor and the strain level in our sample is difficult. It is proposed that upon uniaxial tensile strain, the valence band edge would be shifted downward,[38] which would induce the Fermi energy closer to the band edge, leading to lower carrier density and thus higher resistance. Such scenario is verified by the conventional Hall measurements. As shown in Fig. 2(c), the linear Hall resistance is measured under different status of strain, indicating the Fermi energy lies in the valence band with hole as dominated carrier. With increasing the tensile strain by tuning $E_P$ from $-15$ to $15$ kV/cm, the carrier density is reduced from $6.6 \times 10^{13}$ to $5 \times 10^{13}$ cm$^{-2}$. The change of resistance of WSe$_2$ under various temperature is plotted in Fig. 2(d), which is quite similar to that of gauge (inset of Fig. 2(d)), indicating the strain origins from PMN-PT substrate.

Nonlinear transport is investigated under different strains and temperatures. Figure 3(a) shows the first-harmonic current-voltage characteristics. Linear behaviors are observed, indicating the ohmic contact. Since nonlinear Hall effect is described as the second-order response to electric field $E$, an a.c. current with frequency $\omega$ would then induce a Hall voltage with double frequency $2\omega$. Phenomenologically, the nonlinear Hall effect is understood in terms of orbital magnetization induced anomalous Hall effect. The Berry curvature dipole $D$ is proposed to result in current-induced orbital magnetization $M \propto D \cdot E$,[30] which further induces an anomalous Hall effect in a second-order response to $E$, *i.e.*, the nonlinear Hall effect, as shown in the insets of Figs. 3(b) and 3(c). Additionally, the $\mathbf{D} \cdot \mathbf{E}$ term implies the nonlinear Hall voltage is nonzero only when the electric field has nonzero component along the Berry curvature dipole direction, which is the zigzag orientation in strained TMDs.[11] Here the piezoelectric field $E_P$ is applied along the zigzag direction confirmed by SHG measurements. The applied a.c. current $I$ is aligned with $E_P$ and thus zigzag direction. The second-harmonic Hall voltage $V_{xy}^{2\omega}$ was measured under different strains at 140 K as plotted in Fig. 3(b). The measurement configuration is shown in the inset. The $V_{xy}^{2\omega}$ demonstrates quadric dependence on electric field and also strain tunable. Upon increasing $E_P$, the $V_{xy}^{2\omega}$ is significantly enhanced. When the current polarity is reversed, the current-induced orbital magnetization is also reversed (inset in Fig. 3(c)). The sign change of both current and orbital magnetization will induce the polarity of nonlinear Hall current to be unchanged. As shown in Fig. 3(c), when the current source-drain terminals and the Hall voltage detection terminals are exchanged, the measured $V_{xy}^{2\omega}$ changes sign, which is distinct from the conventional Hall effect. For clarity, the $V_{xy}^{2\omega}$ as a function of $I^2$ are plotted in Fig. 3(d), where linear dependences are observed as expected. Further, the second-harmonic longitudinal voltage $V_{xx}^{2\omega}$ is also measured as shown in Fig. 3(e). In contrast to the nonlinear Hall voltage, the $V_{xx}^{2\omega}$ is not observable even under $E_P = 15$ kV/cm.

From the nonlinear Hall measurements, the Berry curvature dipole is estimated by the formula $D = \frac{2\hbar^2 \sigma^3 W}{e^3 \tau} \frac{V_{xy}^{2\omega}}{(I)^2}$,[18] where $\sigma$ is the longitudinal conductivity, $\tau$ is the scattering time, and $W$ is the channel width. The scattering time $\tau \sim 1.2 \times 10^{-14}$ s is calculated through $\sigma = \frac{ne^2\tau}{m^*}$ with effective mass $m^* = 0.54 m_e$[42] and carrier density estimated from the conventional Hall measurements. The calculated Berry curvature dipole is shown in Fig. 3(f). It clearly shows that the Berry curvature dipole of WSe$_2$ is highly tunable by tuning strain (inset of Fig. 3(f)) and significantly enhanced when increasing $E_P$. Further, the Berry curvature dipole versus $E_P$ at various temperatures is also shown in Fig. 3(f). The Berry curvature dipole is decreased with decreasing temperature, which is attributed to the suppression of piezoelectric coefficient of PMN-PT crystal at low temperatures.[40] In addition to the Berry curvature dipole, the corresponding current-induced orbital magnetization per current density $M/J$, is also shown in Fig. 3(f), which can reach as large as 60. The $M/J$ is estimated based on the formula $M = -\frac{e^2\tau}{2t\hbar^2}\Delta'(\boldsymbol{D}\cdot\boldsymbol{E})$,[30] where $t$ and $\Delta'$ are the thickness and energy gap of monolayer WSe$_2$, respectively. In contrast to the in-plane spin polarization induced by the Rashba or topological surface state-Edelstein effect, the orbital magnetization is along out-of-plane direction, promising for high-efficient perpendicular magnetization switching. It is worth noting that the measured Berry curvature dipole $D \sim 4$ nm in the strained WSe$_2$ seems larger than the theoretical predictions.[11] This difference may be due to the enhanced Rashba spin-orbit coupling caused by the large interfacial electric field as gating the sample by ion-liquid, thus strengthening the Berry curvature near the valance band edge. Moreover, despite the different structures, the calculated dipole in our case has the same order of magnitude with that observed in WTe$_2$.[18]

Figure 4 shows the nonlinear Hall effect under $E_P = 15$ kV/cm at various temperatures. The nonlinear Hall signals show non-monotonic dependence on temperature. Upon decreasing temperature, the nonlinear Hall signals first increase and then decrease. The non-monotonic dependence could be understood in terms of competition between temperature-dependent strain of PMN-PT crystals and thermal fluctuations. At high temperatures, the strain is large, inducing large Berry curvature dipole, while the strong thermal fluctuations would smear the nonlinear Hall signals. By comparison, at low temperatures, the thermal fluctuations are suppressed, but the strain is also decreased. Thus, the maximum nonlinear Hall signals emerge at moderate temperature ~80 K in our experiments.

In summary, we demonstrate the strain-controlled Berry curvature dipole, orbital magnetization and nonlinear Hall effect in monolayer WSe$_2$. Upon applying different strains through PMN-PT substrate, the Berry curvature dipole is highly tunable by electrical means. The emergent nonlinear Hall effect could be exploited to detect the Berry curvature distribution near band edge,[16] helping for deeper understanding about Berry curvature-related physics. Moreover, the Berry curvature dipole is directly related to the current-induced orbital magnetization in TMDs.[29] In contrast to the in-plane spin polarization from spin-orbit coupling effect, the orbital magnetization is along out-of-plane direction, which is significant for the perpendicular magnetization switching by "orbit-magnetism torque" (OMT).

**ACKNOWLEDGEMENTS**

This work was supported by National Key Research and Development Program of China (Nos. 2018YFA0703703 and 2016YFA0300802), and NSFC (Nos. 91964201, 61825401, and 11774004).

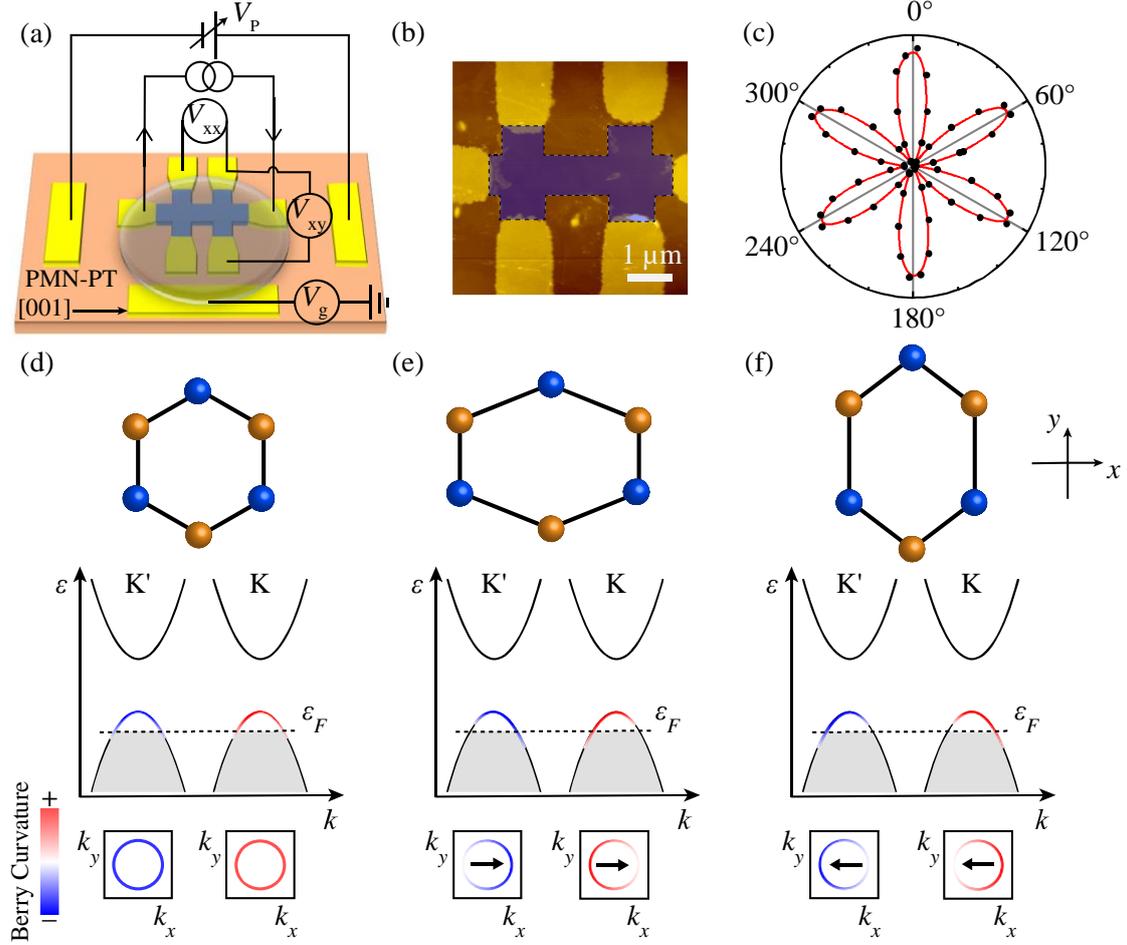

Fig. 1. Device characterizations and illustration of the nonzero Berry curvature dipole induced by uniaxial strain. (a) Schematic illustration of the monolayer WSe$_2$ device. The voltage $V_P$ is applied on the PMN-PT substrate to generate piezoelectric field. (b) The atomic force microscope image of a typical device. (c) The polarization-resolved second harmonic generation of the device in (b), where the directions with maximum intensity correspond to armchair direction of monolayer WSe$_2$. (d-f) Illustrations of nonzero Berry curvature dipole induced by uniaxial strain along zigzag direction ($x$ direction). The top panels exhibit the WSe$_2$ crystal under various strain, where yellow denotes wolfram and blue denotes selenium. The middle panels show the band structure under various strain. The color is used to describe the strength of Berry curvature near the valence band edge. The bottom panels show the Berry curvature distributions on Fermi surface. The black arrows denote the direction of Berry curvature dipole.

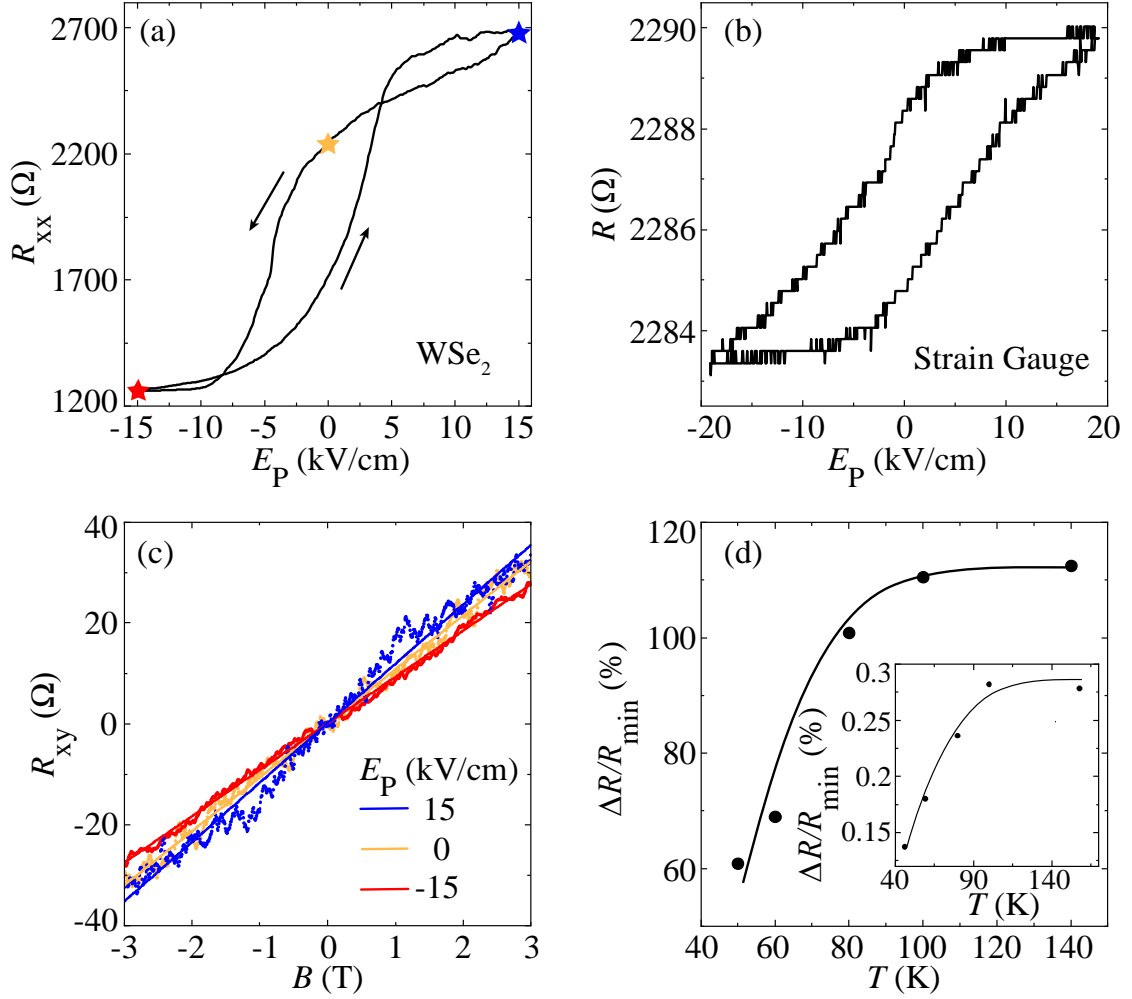

Fig. 2. Strain effect on transport properties of monolayer WSe$_2$. (a) The longitudinal resistance of monolayer WSe$_2$ as a function of piezoelectric field $E_P$ at 140 K. (b) The resistance of Au strain gauge, as a function of $E_P$ at 140 K. (c) The Hall resistance of monolayer WSe$_2$ under various strains as marked by the pentagrams in (a) with the same color. (d) The change of longitudinal resistance in a hysteresis loop as sweeping $E_P$, defined as $\frac{\Delta R}{R_{min}} = \frac{R_{max}-R_{min}}{R_{min}} \times 100\%$, as a function of temperature. The inset shows the change of gauge resistance versus temperature.

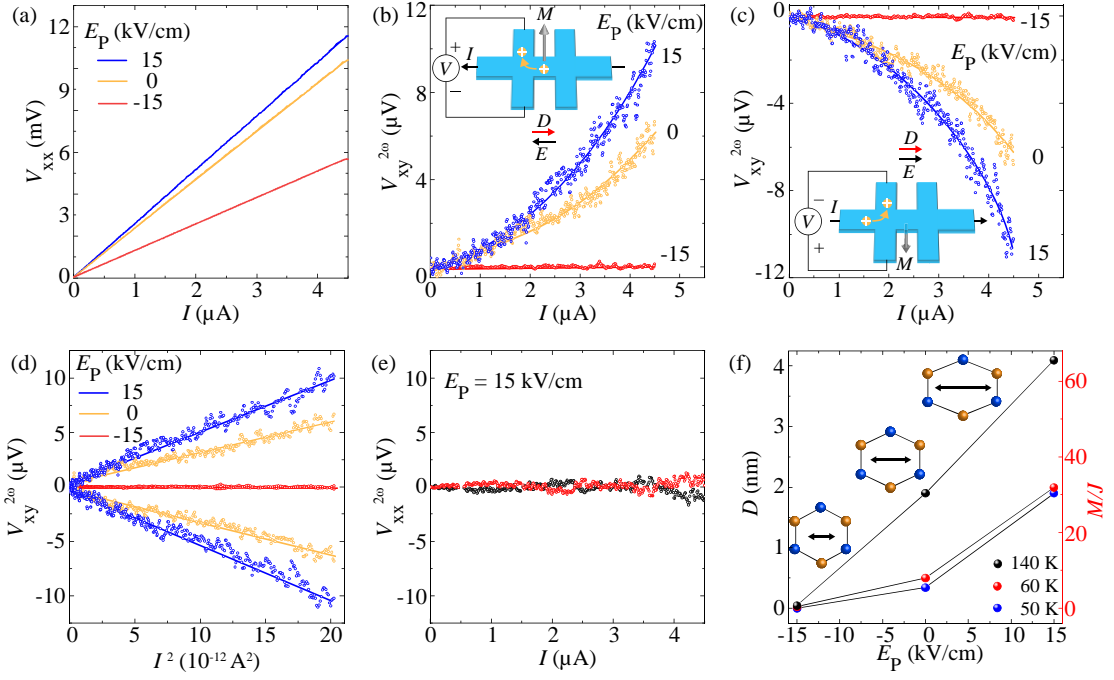

Fig. 3. Strain tunable nonlinear Hall effect and Berry curvature dipole. (a) The longitudinal voltage $V_{xx}$ versus the applied current under various strains at 140 K. (b) and (c) The second-harmonic Hall voltage $V_{xy}^{2\omega}$ versus current $I$ at 140 K with different measurement configurations shown in the insets, respectively. When the applied electric filed $E$ is antiparallel (parallel) with the Berry curvature dipole $D$, an upward (downward) out-of-plane orbital magnetization $M$ is generated, which further induces the nonlinear Hall effect. (d) The $V_{xy}^{2\omega}$ versus $I^2$ at 140 K. The positive and negative $V_{xy}^{2\omega}$ corresponds to the results in (b) and (c), respectively. (e) The second-harmonic longitudinal voltage $V_{xx}^{2\omega}$ versus $I$ at 140 K. (f) The estimated Berry curvature dipole $D$ and orbital magnetization per current density $M/J$ versus piezoelectric field $E_P$ at various temperatures.

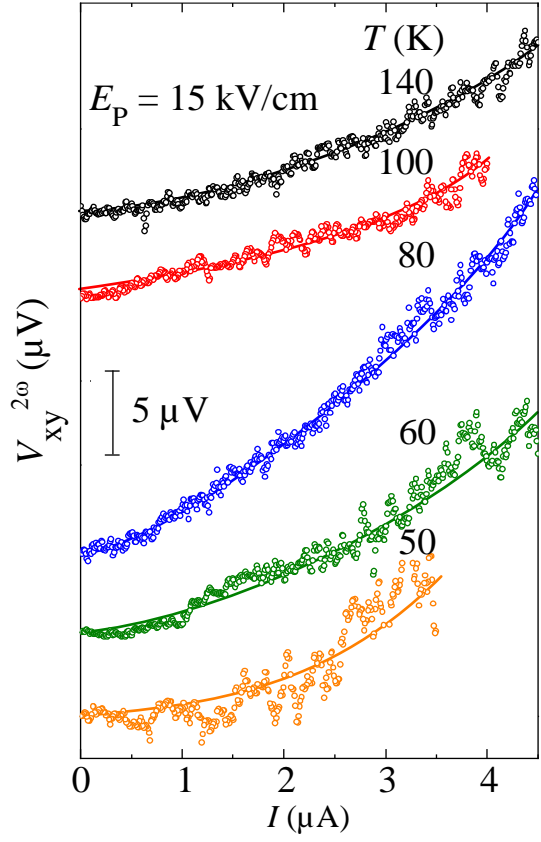

Fig. 4. The second-harmonic Hall voltage $V_{xy}^{2\omega}$ versus $I$ at various temperatures under $E_P = 15$ kV/cm. The curves are shifted for clarity.